\documentclass{article}
\usepackage{frascatiphys_db,here,graphicx,subfigure}
\begin{document}
\title{ 
Study of scalar mesons in chiral Lagrangian frameworks
}
\author{
Deirdre Black and Jonathan Gaunt       \\
{\em Department of Physics, Cavendish Laboratory, Cambridge CB3 OHE, UK}
}
\maketitle
\baselineskip=11.6pt
\begin{abstract}
We review two previous approaches to studying pseudoscalar meson-meson scattering amplitudes to beyond 1 GeV using non-linear and linear chiral Lagrangians.  In these approaches we use two different unitarisation techniques - a generalised Breit Wigner prescription and K-matrix unitarization respectively.  We also report some preliminary findings on K-matrix unitarisation of the I=J=0 $\pi\pi$ scattering in a non-linear chiral Lagrangian approach and make some remarks about the light scalar mesons.    
\end{abstract}
\baselineskip=14pt
\section{Introduction}
Pseudoscalar meson-meson scattering up to the 1-2GeV energy range is of interest for 
several related reasons.  On the one hand this region is beyond that where chiral perturbation theory has
traditionally been applied and below that where we can use perturbative QCD, so it is a challenge to 
develop a framework to calculate these amplitudes from first principles.  At the same time 
there are many resonances in this region, some of which are controversial from the point of view of 
establishing their properties experimentally and their quark substructure.  In particular, the scalar 
mesons are a long-standing puzzle in meson spectroscopy because, for example, there are too many states to fit
into a single SU(3) nonet and the masses and decay patterns of some of the scalar resonances are not what would would 
expect for quark-antiquark scalar states.  This talk is based on approaches developed by the Syracuse group.  Many 
other interesting approaches are given in the proceedings of this conference and also cited in the references given in 
the bibliography.

\section{Non-linear chiral Lagrangian approach to meson-meson scattering}

We begin\cite{pipi_nlsm},\cite{pik_nlsm} with the conventional chiral Lagrangian, including
only pseudoscalars:
\begin{equation}
{\cal {L}}_1=
- { {F_\pi^2} \over 8} {\rm Tr} 
\left( \partial_\mu U \partial_\mu U^\dagger  \right)
+{\rm Tr}
\left[
       {\cal B} \left( U + U^\dagger \right)
\right],
\label{Lps}
\end{equation}
in which $\displaystyle{U=e^{2i{\phi\over F_\pi}}}$, with $\phi$ the
$3 \times 3$ matrix of pseudoscalar fields and $F_\pi = 132 $ MeV the
pion decay constant. ${\cal B}$ is a diagonal matrix $(B_1, B_1, B_3)$
with $B_1=m_\pi^2 F_\pi^2/8=B_2$ and $B_3=F_\pi^2(m_K^2-m_\pi^2/2)/4$. 

We add a nonet of scalar mesons, which transform like external fields under chiral transformations.  It turns 
out\cite{putative} that the trilinear scalar-pseudoscalar-pseudoscalar interaction that follows from the general 
chiral invariant extension of ${\cal L}_1$ to include a scalar meson nonet is given by  
\begin{eqnarray}
{\cal L}_{N\phi \phi} &=&
A{\epsilon}^{abc}{\epsilon}_{def}N_{a}^{d}{\partial_\mu}{\phi}_{b}^{e}{\partial_\mu}{\phi}_{c}^{f}
+ B {\rm Tr} \left( N \right) {\rm Tr} \left({\partial_\mu}\phi
{\partial_\mu}\phi \right) \nonumber \\
&+& C {\rm Tr} \left( N {\partial_\mu}\phi \right) {\rm
Tr} \left( {\partial_\mu}\phi \right) 
 + D {\rm Tr} \left( N \right) {\rm Tr}
\left({\partial_\mu}\phi \right)  {\rm Tr} \left( {\partial_\mu}\phi \right)
\label{interactions}
\end{eqnarray}
The first term of (\ref{interactions}) may be eliminated in favor of the more standard form ${\rm Tr} \left(
N{\partial_\mu}{\phi}{\partial_\mu}{\phi}\right)$, but is interesting because it is the OZI rule conserving term 
for a dual diquark-antidiquark type nonet mentioned below.

The scalar particles with non-trivial quantum numbers are given by:
\begin{equation}
N = \left[ \begin{array}{c c c}
N_1^1&a_0^+&\kappa ^+\\
a_0^-&N_2^2&\kappa ^0\\
\kappa^-&{\bar \kappa}^0&N_3^3
\end{array} \right]
\end{equation}
with $a_0^0 = (N_1^1 - N_2^2)/ \sqrt 2$.  There are
two iso-singlet states: the combination $(N_1^1 + N_2^2
+ N_3^3)/ \sqrt 3$ is an $SU(3)$ singlet while $(N_1^1
+ N_2^2 - 2N_3^3)/\sqrt 6$ belongs to an $SU(3)$ octet.  These will in
general mix with each other when $SU(3)$ is broken.  We can write the general 
mass term\cite{putative}
\begin{equation}
{\cal L}_{mass} = -a {\rm Tr}(NN) - b {\rm Tr}(NN{\cal M}) - c {\rm
Tr}(N)Tr(N) - d {\rm Tr}(N) {\rm Tr}(N{\cal M}),
\label{mixing-mass-Lag}
\end{equation}
where a, b, c and d are real constants. ${\cal M}$ is the ``spurion matrix''
${\cal M}={\rm diag}(1,1,x)$ , $x$ being the ratio of strange to non-strange quark
masses in the usual interpretation.

We take a convention where the physical particles, $\sigma$ and $f_0$, which diagonalize
the mass matrix obtained from (\ref{mixing-mass-Lag}) are related to the basis states $N_3^3$ and $(N_1^1 +
N_2^2)/{\sqrt 2}$ by
\begin{equation}
\left( \begin{array}{c} \sigma\\ f_0 \end{array} \right) = \left(
\begin{array}{c c} {\rm cos} \theta_s & -{\rm sin} \theta_s \\ {\rm sin}
\theta_s & {\rm cos} \theta_s \end{array} \right) \left( \begin{array}{c}
N_3^3 \\ \frac {N_1^1 + N_2^2}{\sqrt 2} \end{array} \right),
\label{mixing-convention1}
\end{equation}
For a given set of inputs for the masses of the four scalar mesons $\sigma$, $f_0(980)$, $a_0(980)$ and $\kappa$ the constants a, b, c and d are fixed and there are two possible solutions for the mixing angle $\theta_s$.   

\begin{figure}
    \begin{center}
        {\includegraphics[scale=0.35,angle = 270]{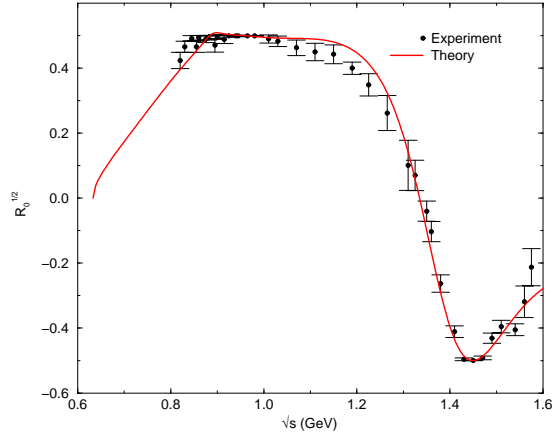}}
        \caption{\it Best fit to the experimental data\cite{pikdata} for the Real part of the I=$\frac{1}{2}$, J=0, $\pi$-K scattering amplitude in our non-linear chiral Lagrangian model\cite{pik_nlsm} with generalised Breit-Wigner prescription. }
\label{pikfit}
    \end{center}
\end{figure}

We note that there are different possibilities, in addition to quark-antiquark configurations, for the underlying quark substructure of $N$ which all give rise to the same SU(3) transformation properties.  For example, forming diquark objects
\begin{equation}
T_a = \epsilon _{abc} {\bar q}^b {\bar q}^c , \quad \quad {\bar T}^a =
\epsilon^{abc}q_{b}q_{c},
\label{dual-quarks}
\end{equation}
where the antisymmeterisation of the quark fields is implicit, we can form a pure tetraquark scalar nonet as follows:   
\begin{equation}
N_a^b \sim {T_a}{{\bar T}^b} \sim 
\left[ \begin{array}{c c c}
\bar {s} \bar {d} ds &\bar {s} \bar {d} us&\bar {s} \bar {d} ud\\
\bar {s} \bar {u} ds&\bar {s} \bar {u} us&\bar {s} \bar {u} ud\\
\bar {u} \bar {d} ds&\bar {u} \bar {d} us& \bar {u} \bar {d} ud
\end{array} \right]
\label{multiquark-nonet}
\end{equation}
or construct linear combinations of $q \bar q$ and $qq \bar q \bar q$ nonets. 
We studied s-wave pseudoscalar meson scattering in a framework beginning with Eqs.(\ref{Lps}) and (\ref{interactions}).  If we begin with the tree-level scattering amplitudes, which due to chiral symmetry give good agreement with experiment close to the scattering threshold, we find that they soon deviate from the experimental data.  They also violate unitarity.  The approach that we took was to add an imaginary piece by hand to the tree-level propogator of the s-channel resonance.  For $\pi$ K scattering we called the lightest strange scalar resonance $\kappa$ and made the substitution:
\begin{equation}
{m_{\kappa}^2 - s} \longrightarrow {m_{\kappa}^2 - s -
im_{\kappa}G'_{\kappa}}
\label{GBW}
\end{equation}
in the denominator of the s-channel s-wave amplitude.  In order to fit to experiment the quantity $G^\prime_\kappa$ was left as a free parameter, not necessarily equal to the perturbative width, $G_{\kappa}$ say.  Our fit\cite{pik_nlsm}, shown in Fig. \ref{pikfit} gave $\displaystyle{\frac
{G_{\kappa}}{G'_\kappa} = 0.13}$ showing a substantial deviation from a Breit-Wigner resonance for which this ratio would be exactly equal to 1. Good agreement with experiment was also found\cite{pipi_nlsm} with this generalised Breit-Wigner prescription for the case of $\pi\pi$ scattering.  The other fitting parameters are the scalar-pseudoscalar-pseudoscalar coupling constants, which can all be written in terms of the four coefficients in the interaction terms in Eq. (\ref{interactions}), the scalar meson masses and mixing angle. 

We note also that, in addition to neatly explaining the mass ordering and general pattern of decays of the scalar states below 1 GeV, a multiquark interpretation for these states is also suggested by the value of the scalar meson octet-singlet mixing angle $\theta_s$ defined in Eq. (\ref{mixing-convention1}), which was a parameter fixed by our fits.  Our best fit was about $-20^o$ which, in our mixing convention, would be close to ideal mixing for a ``dual'' diquark-antidiquark nonet.

\section{Pseudoscalar meson-meson scattering in Linear Sigma Models}

In the three flavor linear sigma model the pseudoscalar and scalar mesons appear 
together since the model is constructed from the $3 \times
3$ matrix field
\begin{equation}
M=S+i\phi,
\end{equation}
where $S=S^\dagger$ represents a scalar nonet and $\phi= \phi^\dagger$
a pseudoscalar nonet.  Under a chiral transformation $q_L \rightarrow
U_L q_L$, $q_R \rightarrow U_R q_R$ of the fundamental left and right
handed light quark fields, M is defined to transform as 
\begin{equation}  
M \longrightarrow U_L M U_R^\dagger.
\end{equation}
We considered a general non-renormalizable Lagrangian\footnote{See \cite{lsm} and references therein for more detail} of the form
\begin{equation}
{\cal L} = - \frac{1}{2} {\rm Tr} \left( \partial_\mu \phi \partial_\mu \phi
\right) - \frac{1}{2} {\rm Tr} \left( \partial_\mu S \partial_\mu S 
\right) - V_0 - V_{SB},
\label{LsMLag}
\end{equation}
where $V_0$ is an arbitrary function of the independent $SU(3)_L
\times SU(3)_R \times U(1)_V$ invariants ${\rm Tr} \left( M M^\dagger \right)$, ${\rm Tr} \left( M
M^\dagger M M^\dagger \right)$, ${\rm Tr} \left( (M
M^\dagger)^3 \right)$ $6\left( {\rm det} M + {\rm det}
M^\dagger \right)$.  Of these, only $I_4$ is not invariant under $U(1)_A$.  The symmetry
breaker $V_{SB}$ has the minimal form 
\begin{equation}
V_{SB} = -2 \left( A_1 S_1^1 + A_2 S_2^2 + A_3 S_3^3 \right),
\end{equation}
where the $A_a$ are real numbers which turn out to be proportional to
the three light (``current'' type) quark masses. 
In this model there are many constraints among the parameters.  For example, many of the trilinear scalar-pseudoscalar-pseudoscalar coupling constants are predicted in terms of the pseudoscalar and scalar meson masses.  Another difference is that [compare with Eq. (\ref{interactions})] this trilinear interaction does not involve derivatives.  Both models give the ``current algebra'' results in the limit where the scalar mesons are integrated out.  

\begin{figure}
    \begin{center}
        {\includegraphics[scale=0.35, angle=270]{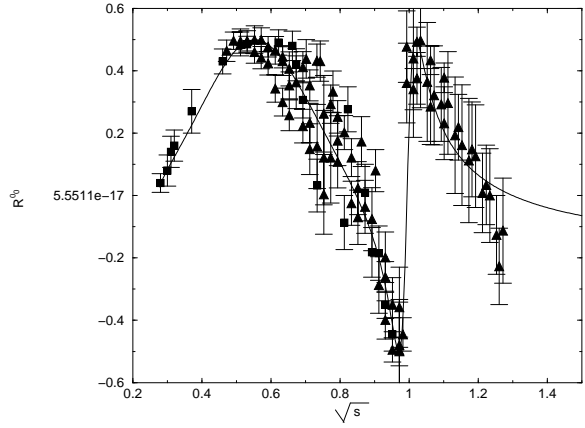}}
        \caption{\it Best fit to the experimental data\cite{pipidata} for the Real part of the I=J=0 $\pi\pi$ scattering amplitude in a linear sigma model with K-matrix unitarisation\cite{lsm}.}
\label{pipifit}
    \end{center}
\end{figure}

If we calculate the tree level s-wave amplitudes they deviate from experiment and also violate unitarity as we go beyond the threshold region.  We used\cite{lsm} the well-known K-matrix procedure to unitarise the linear sigma model amplitudes and then checked if the resulting unitary amplitudes can give a good fit to data.  In the standard parameterization \cite{chung} of a given partial wave S-matrix:
\begin{equation}
S = \frac {1 + iK}{1 - iK} \equiv 1 + 2iT,
\label{regularization}
\end{equation}
we identify 
\begin{equation}
K=T_{\rm{tree}}.
\label{Kdef}
\end{equation}
$T_{\rm{tree}}$ is the given partial wave T-matrix computed at tree
level in the Linear Sigma Model and is purely real.  This scheme gives exact unitarity for T
but violates the crossing symmetry which $T_{\rm{tree}}$ itself obeys.  In Fig. \ref{pipifit} we show our best fit to the I=J=0 $\pi\pi$ scattering data.  The parameters in this fit are the ``bare'' masses of the two I=0 scalar mesons in $M$ and their mixing angle. Using these parameters we can solve for the poles in the unitarised amplitude in the complex s plane.  Labelling these poles $z_\sigma$ and $z_{\sigma^\prime}$ we can identify the physical masses and widths as usual from the Real and Imaginary parts, for example $z_\sigma = m_\sigma^2 - im_\sigma\Gamma_\sigma$.

\section{Summary and comparison between models}

We have found good agreement with scattering data in the approaches based on the non-linear chiral Lagrangians outlined in Sections 2 and 3.  We are currently studying scattering using the non-linear chiral Lagrangian approach outlined of Section 2, but employing the K-matrix unitarisation as described in Section 3.  This may make it easier to compare the linear and non-linear chiral Lagrangian models more directly and to understand the effects of the unitarisation prescriptions in themselves.  This was partly motivated by our work on extending the non-linear chiral Lagrangian approach to include vector mesons\cite{VSS}.  This enabled us to study the interesting rare radiative decay processes $\phi \rightarrow \pi \pi \gamma$ and $\phi \rightarrow \pi \eta \gamma$.  We found that the shape of the partial branching fraction depends quite sensitively on whether we use derivative or non-derivative scalar-pseudoscalar-pseudoscalar coupling as in Section 2 or 3 respectively.  

A summary of our results is shown in Table \ref{comparison} for the case of $\pi\pi$ scattering.  In the third and fourth columns we show the results of the analyses described in sections 3 and 2 respectively.  In the fourth column we give the results with and without the inclusion of the $\rho$ vector meson.  In column 2 we show the results of our current analysis, which are preliminary.  However we can see some trends, namely that the $f_0(980)$ parameters are quite stable, whereas the $\sigma$ parameters seem to depend more on the model and, even more, on the unitarisation procedure.  These results are preliminary because we have only done a fit of $\pi\pi$ scattering data over a limited energy range.  Also we have not included the inelastic channel and so the important $K \bar K$ threshold region.  These and a similar study of related scattering channels are interesting directions for future work.

\begin{table}[t]
\centering
\caption{ \it Results for physical I=0 scalar meson parameters from fits to $\pi\pi$ scattering.  Comparison between linear and non-linear chiral Lagrangian models using K-matrix and Generalised Breit-Wigner approaches to unitarisation.
}
\vskip 0.1 in
\begin{tabular}{|l|c|c|c|} \hline
Scalar           &  Non-linear chiral  & SU(3) Linear & Non-linear chiral   \\
parameters &  approach & Sigma Model\cite{lsm} & approach\cite{pipi_nlsm} \\
 (MeV) &  K-matrix  & K-matrix & Generalised B-W \\
 & (preliminary)& & without/with $\rho$ meson \\

\hline
\hline
 $m_\sigma$   & 444 & 457 & 378/559                  \\
\hline 
$\Gamma_\sigma$   & 604 & 632 & 836/370 \\
\hline
$m_f$  & 986 & 993 & 987  \\
\hline
$\Gamma_f$ & 52 & 51 & 65 \\
\hline
\end{tabular}
\label{comparison}
\end{table}
%
%

\section{Acknowledgements}
DB would like to thank the conference organisers for a friendly and stimulating meeting.  
The work in much of this presentation is part of an ongoing collaboration with 
Abdou Abdel-Rehim, Amir Fariborz, Masayasu Harada and Joseph Schechter.  DB is supported
by the Royal Society and JG was supported by a Summer Research Studentship from
Trinity College, Cambridge.

\end{document}